\documentclass[twoside]{ilcws08}
\usepackage[latin1]{inputenc}
\usepackage[dvips]{graphicx,epsfig,color}
\usepackage{wrapfig,rotating}
\usepackage{amssymb,amsmath,array}

\newcommand{\zh}{$Z_{\mathrm{H}}$}
\newcommand{\ah}{$A_{\mathrm{H}}$}
\newcommand{\wh}{$W_{\mathrm{H}}$}

\newcommand{\zhahahahbb}{$A_{\mathrm{H}}Z_{\mathrm{H}} \to A_{\mathrm{H}} A_{\mathrm{H}} bb$}
\newcommand{\nnhnnbb}{$\nu \nu h \to \nu \nu bb$}
\newcommand{\zhnnbb}{$Zh \to \nu \nu bb$}
\newcommand{\zznnbb}{$ZZ \to \nu \nu bb$}
\newcommand{\nnznnbb}{$\nu \nu Z \to \nu \nu bb$}
\newcommand{\gzgbb}{$\gamma Z \to \gamma bb$}

\newcommand{\pt}{$p_{\mathrm{T}}$}

\begin{document}
\title{
%%%%   Paper title goes here  %%%%%%%%%%%%%%
Simulation study of $Z_{\mathrm{H}}A_{\mathrm{H}}$ mode in Littlest Higgs Model with T-Parity} %%
%\LaTeX Template for LCWS08/ILC08 Proceedings} %% 
%***********************************************************************
% AUTHORS INFORMATION AREA
%***********************************************************************
\author{%First Author$^1$ and Second Author$^2$
    Tomonori Kusano$^{1}$,
    Eri Asakawa$^{2}$,
    Masaki Asano$^{3}$,
     Keisuke Fujii$^{4}$,
     Shigeki Matsumoto$^{5}$,\\
    Rei Sasaki$^{1}$,
     Yosuke Takubo$^{1}$
    and
     Hitoshi Yamamoto$^{1}$
% Optional short acknowledgment: remove next line if non-needed
%\thanks{This is an optional funding source acknowledgment.}
% DO NOT MODIFY THE FOLLOWING '\vspace' ARGUMENT
\vspace{.3cm}\\
% Addresses and institutions (remove "1- " in case of a single institution)
%1- First Author's Institution - Department \\
%Address of First Author's Institution - Country
  $^{1}${Department of Physics, Tohoku University, Sendai, Japan} \vspace{.1cm}\\
  	$^{2}${Institute of Physics, Meiji Gakuin University, Yokohama, Japan} \vspace{.1cm}\\
     $^{3}${Institute for Cosmic Ray Research (ICRR), University of Tokyo, Kashiwa, Japan} \vspace{.1cm}\\
     $^{4}${High Energy Accelerator Research Organization (KEK), Tsukuba, Japan} \vspace{.1cm}\\
     $^{5}${Department of Physics, University of Toyama, Toyama, Japan} \\
%% Remove the next three lines in case of a single institution
\vspace{.1cm}\\
%2- Second Author's Institution - Department \\
%Address of Second Author's Institution - Country\\
}
%%***********************************************************************
% END OF AUTHORS INFORMATION AREA
%***********************************************************************

\maketitle  
\begin{abstract}
We investigate a possibility to measure parameters of the Littlest Higgs model with T-parity at the ILC with $\sqrt{s} = 500$ GeV. The model predicts new gauge bosons (\ah, \zh, and \wh), among which the heavy photon (\ah) is a candidate for dark matter. Through Monte Carlo simulations of the process of $e^+ e^- \rightarrow A_{\mathrm{H}} Z_{\mathrm{H}}$, we show that the masses can be determined with accuracies of 16.2\% for $m_{A_{\mathrm{H}}}$ and 4.3\% for $m_{Z_{\mathrm{H}}}$. In additionally, the vacuum expectation value can be determined with accuracies of 4.3\%.
\end{abstract}

\section{Introduction}
The Littlest Higgs Model with T-parity is one of the attractive candidates for physics beyond the Standard Model because it solves both the little hierarchy and dark matter problems simultaneously. One of the important predictions of the model is the existence of new heavy gauge bosons, where they acquire mass terms through the breaking of the global symmetry necessarily imposed on the model.
The precise measurements of their masses allow us to determine the most important parameter of the model, namely the vacuum expectation value of the breaking. 

At the Large Hadron Collider (LHC), it is difficult to determine the properties of heavy gauge bosons, because they have no color charge \cite{Cao:2007pv}. On the other hand, the ILC will provide an ideal environment to measure the properties of heavy gauge bosons. Heavy gauge bosons are expected to be produced in a clean environment at the ILC, which enables us to determine their properties precisely. We study the sensitivity of the measurements to the masses of heavy gauge bosons and the vacuum expectation value at the ILC based on a realistic Monte Carlo simulation.

\section{Simulation framework}
\label{sec:mctool}
In order to perform a numerical simulation, we need to choose a representative point in the parameter space of the Littlest Higgs model with T-parity. Firstly, the model parameters should satisfy the experimental results of the precise measurements on the electroweak parameters \cite{Yao:2006px}. In addition, the cosmological observation of dark matter relics also gives important information. For that reason, we consider not only the electroweak precision measurements but also the WMAP observation \cite{Komatsu:2008hk} to choose a point in the parameter space. 
At selected representative point, the vacuum expectation value($f$) is 580 GeV, and the masses of the heavy gauge boson are 81.9 GeV and 369 GeV for $m_{A_{\mathrm{H}}}$ and $m_{Z_{\mathrm{H}}}$, respectively.

\begin{wrapfigure}{r}{0.3\columnwidth}
\centerline{\includegraphics[width=0.25\columnwidth]{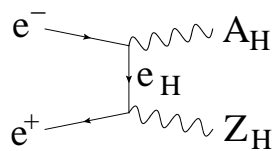}}
  \caption{\small Diagram for signal process.}
  % $e^+e^- \rightarrow A_{\mathrm{H}}Z_{\mathrm{H}}$ }
  \label{fig:signal diagrams}
\end{wrapfigure}

Because sum of $m_{A_{\mathrm{H}}}$ and $m_{Z_{\mathrm{H}}}$ is less than 500 GeV,  $m_{A_{\mathrm{H}}}$ and $m_{Z_{\mathrm{H}}}$ can be generated through the process $e^+e^- \rightarrow A_{\mathrm{H}}Z_{\mathrm{H}}$ at $\sqrt{s} = 500$ GeV. 
Feynman diagram for the signal process is shown in Fig. \ref{fig:signal diagrams}. \zh~decays into $A_{\mathrm{H}} h$ with almost 100\% branching fractions.

We have used MadGraph \cite{madgraph} to generate signal events of the Little Higgs model, while Standard Model events have been generated by Physsim \cite{physsim}. Initial-state radiation and beamstrahlung have not been included in the event generations. We have ignored the finite crossing angle between the electron and positron beams. In both event generations, helicity amplitudes were calculated using the HELAS library \cite{helas}, which allows us to deal with the effect of gauge boson polarizations properly. Phase space integration and the generation of parton 4-momenta have been performed by BASES/SPRING \cite{bases}. Parton showering and hadronization have been carried out by using PYTHIA6.4 \cite{pythia}, where final-state tau leptons are decayed by TAUOLA \cite{tauola} in order to handle their polarizations correctly.

The generated Monte Carlo events have been passed to a detector simulator called JSFQuickSimulator, which implements the GLD geometry and other detector-performance related parameters \cite{glddod}. %In the detector simulator, hits by charged particles at the vertex detector and track parameters at the central tracker are smeared according to their position resolutions, taking into account correlations due to off-diagonal elements in the error matrix. Since calorimeter signals are simulated in individual segments, a realistic simulation of cluster overlapping is possible. Track-cluster matching is performed for the hit clusters in the calorimeter in order to achieve the best energy flow measurements. 
%The resultant detector performance in our simulation study is summarized in Table \ref{tb:GLD}.

\section{Analysis} \label{sec:analysis}
The simulation has been performed at $\sqrt{s} =$ 500 GeV with an integrated luminosity of 500 fb$^{-1}$. The heavy gauge bosons, \ah~and \zh~, are produced with the cross section of 1.9 fb. Since \zh~decays into \ah~and the Higgs boson, the signature is a single Higgs boson in the final state, mainly 2 jets from $h \to b\bar{b}$ (with a 55\% branching ratio). We, therefore, define \zhahahahbb~as our signal event. For background events, contribution from light quarks was not taken into account because such events can be rejected to negligible level after requiring the existence of two $b$-jets, assuming a $b$-tagging efficiency of 80\% for $b$-jets with 15\% probability to misidentify a $c$-jet as a $b$-jet. This $b$-tagging performance was estimated by the full simulation assuming a typical ILC detector. Signal and background processes considered in this analysis are summarized in Table \ref{tb:zhevlst}. %Figure \ref{fig:evtdsp} shows a typical \zhah~event as seen in the detector simulator.

%\begin{wraptable}{l}{0.5\columnwidth}
  \begin{table}[t]
  \center{
  \begin{tabular}{l|r|r|r}
   \hline
   Process  & Cross sec. [fb] & \multicolumn{2}{|c}{\# of events } \\
   \cline{3-4}
   &  &generated & after all cuts \\
   \hline
   \zhahahahbb& 1.05            & 525          & 272             \\
   \nnhnnbb & 34.0            & 17,000       & 3,359           \\
   \zhnnbb  & 5.57            & 2,785        & 1,406           \\
   $tt \to WWbb$     & 496             & 248,000      & 264             \\
   \zznnbb  & 25.5            & 12,750       & 178           \\
   \nnznnbb & 44.3            & 22,150       & 167             \\
   \gzgbb   & 1,200           & 600,000      & 45               \\
   \hline
  \end{tabular}
 \caption{\small Signal and backgrounds processes.}
 \label{tb:zhevlst}
 }
\end{table}
%\end{wraptable}

The clusters in the calorimeters are combined to form a jet if the two clusters satisfy $y_{ij} < y_{\mathrm{cut}}$. $y_{ij}$ is defined as
\begin{equation}
 y_{ij} = \frac{2 E_{i} E_{j} (1 - \cos \theta_{ij})}{E_{\mathrm{vis}}^{2}},
\end{equation}
where $\theta_{ij}$ is the angle between momenta of two clusters, $E_{i(j)}$ are their energies, and $E_{\mathrm{vis}}$ is the total visible energy. All events are forced to have two jets by adjusting $y_{\mathrm{cut}}$. We have selected events with the reconstructed Higgs mass in a window of 100-140 GeV. In order to suppress the \nnhnnbb~background, the transverse momentum of the reconstructed Higgs bosons (\pt) is required to be above 80 GeV. This is because the Higgs bosons coming from the $WW$ fusion process, which dominates the \nnhnnbb~background, have \pt~mostly below W mass. Finally, multiplying the efficiency of double $b$-tagging ($0.8 \times 0.8 = 0.64$), we are left with 272 signal and 5,419 background events as shown in Table \ref{tb:zhevlst}, which corresponds to a signal significance of 3.7 ($= 272/\sqrt{5419}$) standard deviations. The indication of the new physics signal can hence be obtained at $\sqrt{s} = 500$ GeV.

The \ah~and \zh~boson masses can be estimated from the edges of the distribution of the reconstructed Higgs boson energies. This is because the maximum and minimum Higgs boson energies ($E_{\mathrm{max}}$ and $E_{\mathrm{min}}$) are written in terms of these masses,
\begin{eqnarray}
 E_{\mathrm{max}}
 &=& 
 \gamma_{Z_{\mathrm{H}}} E^{\ast}_{h}
 + 
 \beta_{Z_{\mathrm{H}}} \gamma_{Z_{\mathrm{H}}} p^{\ast}_{h},
 \nonumber \\ 
 E_{\mathrm{min}}
 &=& 
 \gamma_{Z_{\mathrm{H}}} E^{\ast}_{h}
 - 
 \beta_{Z_{\mathrm{H}}} \gamma_{Z_{\mathrm{H}}} p^{\ast}_{h},  
 \label{eq:eedge}
\end{eqnarray}
where $\beta_{Z_{\mathrm{H}}} (\gamma_{Z_{\mathrm{H}}})$ is the $\beta (\gamma)$ factor of the \zh~boson in the laboratory frame, while $E^{\ast}_{h} (p_{h}^{\ast})$ is the energy (momentum) of the Higgs boson in the rest frame of the \zh~boson. Note that $E^{\ast}_{h}$ is given as $(M_{Z_{\mathrm{H}}}^2 + M_h^2 - M_{A_{\mathrm{H}}}^2)/(2M_{Z_{\mathrm{H}}})$.

\begin{figure}[hbtp]
  \begin{center}
  \includegraphics[width=\hsize]{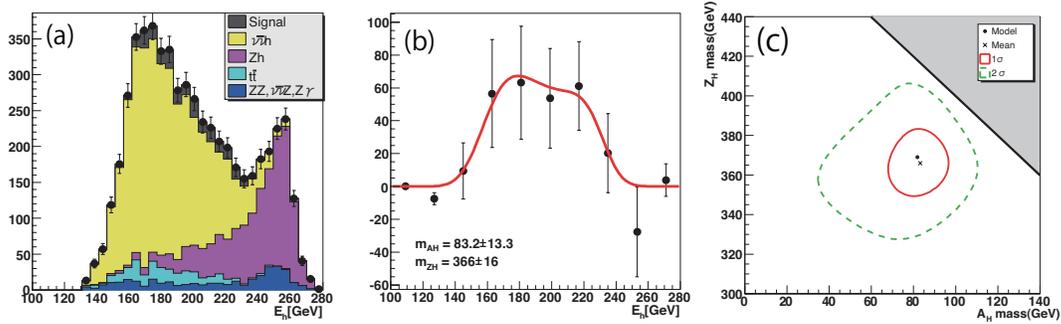}
 \end{center}
 \caption{\small (a) Energy distribution of the reconstructed Higgs bosons with remaining backgrounds after the selection cuts. (b) Energy distribution of the Higgs bosons after subtracting the backgrounds. The distribution is fitted by a line shape function determined with a high statistics signal sample. (c)Probability contour corresponding to 1- and 2-$\sigma$ deviations from the best fit point in the \ah~and \zh~mass plane. The shaded area in shows the unphysical region of $m_{A_{\mathrm{H}}} + m_{Z_{\mathrm{H}}} > 500$ GeV.}
 \label{fig:hene}
\end{figure}

Figure \ref{fig:hene}(a) shows the energy distribution of the reconstructed Higgs bosons with remaining backgrounds. We obtained the signal distribution after subtracting backgrounds as shown in Fig.\ref{fig:hene}(b). The endpoints, $E_{\mathrm{max}}$ and $E_{\mathrm{min}}$, have been derived by fitting the distribution with a line shape function which was determined by a high statistics signal sample. The fit resulted in $m_{A_{\mathrm{H}}}$ and $m_{Z_{\mathrm{H}}}$ being $83.2 \pm 13.3$ GeV and $366.0 \pm 16.0$ GeV, respectively, which should be compared to their true values: 81.85 GeV and 368.2 GeV. Figure \ref{fig:hene}(c) shows the probability contour for the masses of \ah~and \zh~. Since the masses of the heavy gauge bosons depend on the vacuum expectation value, $f$ can be determined by measurement of the gauge boson masses. We obtained  $f = 576.0 \pm 25.0$ GeV. The input value of $f$ is 580 GeV in our simulation study.

\section{Summary}
At the ILC with $\sqrt{s} = 500$ GeV, it is possible to produce \ah~ and \zh~bosons with a signal significance of 3.7-sigma level. By using the energy distribution of the Higgs bosons from the \zh~decays, the masses of these bosons can be determined with accuracies of 16.2\% for $m_{A_{\mathrm{H}}}$ and 4.3\% for $m_{Z_{\mathrm{H}}}$. From the results, it turns out that the vacuum expectation value can be determined with accuracy of 4.3\%. The details on this analysis can be found in \cite{thesis}.

\section{Acknowledgments}

We would like to thank all the members of the ILC physics subgroup \cite{Ref:subgroup} for useful discussions. They are grateful to the Minami-tateya group for the help extended in the early stage of the event generator preparation. This work is supported in part by the Creative Scientific Research Grant (No. 18GS0202) of the Japan Society for Promotion of Science and the JSPS Core University Program.

% ****************************************************************************
% BIBLIOGRAPHY AREA
% ****************************************************************************

\begin{footnotesize}
% IF YOU DO NOT USE BIBTEX, USE THE FOLLOWING SAMPLE SCHEME FOR THE REFERENCES
% ----------------------------------------------------------------------------

% ----------------------------------------------------------------------------

% IF YOU USE BIBTEX,
% - DELETE THE TEXT BETWEEN THE TWO ABOVE DASHED LINES
% - UNCOMMENT THE NEXT TWO LINES AND REPLACE 'Name_Of_Your_BibFile'

%\bibliographystyle{unsrt}
%\bibliography{Name_Of_Your_BibFile}
% example of Name_Of_Your_BibFile.bib
% @Article{Turcato:2006ch,
%      author    = "Turcato, M.",
%  collaboration = "ZEUS and H1",
%      title     = "Lepton flavour violation and charmonium physics at HERA",
%      journal   = "Nucl. Phys. Proc. Suppl.",
%      volume    = "162",
%      year      = "2006", 
%      pages     = "283-287",
%      SLACcitation  = "%%CITATION = NUPHZ,162,283;%%"
% }
% 
% @Unpublished{Gogitidze:2007du,
%      author    = "Gogitidze, N.",
%  collaboration = "H1", 
%      title     = "Prompt photons and particle momentum distributions at
%                   HERA", 
%      year      = "2007",
%      note    = "hep-ex/0701033",
%      SLACcitation  = "%%CITATION = HEP-EX 0701033;%%"
% }

\end{footnotesize}

% ****************************************************************************
% END OF BIBLIOGRAPHY AREA
% ****************************************************************************

\end{document}